\documentclass{ws-ijmpd}
\usepackage[super,compress]{cite}

\usepackage{physics}

\usepackage{tikz}
\usepackage{pgfplots}
\usepackage{tikz-3dplot}
\tdplotsetmaincoords{60}{115}
\pgfplotsset{compat=newest}

\begin{document}

\markboth{T.Ghosh and B.Mukhopadhyay}
{Geometric phase for Dirac Hamiltonian in gravitational fields}

%
\catchline{}{}{}{}{}
%

\title{Geometric phase for Dirac Hamiltonian under gravitational fields in the non-relativistic regime}

\author{Tanuman Ghosh}

\address{Raman Research Institute \\ Bangalore, 560080, India\\
tanuman@rri.res.in}

\author{Banibrata Mukhopadhyay}

\address{Department of Physics, Indian Institute of Science \\ 
Bangalore 560012, India\\
bm@iisc.ac.in}

\maketitle

\begin{history}
\received{Day Month Year}
\revised{Day Month Year}
\end{history}

\begin{abstract}
We show the appearance of geometric phase in a Dirac particle traversing in non-relativistic limit in a time-independent gravitational 
field. This turns out to be similar to the one originally described as a
	geometric phase in magnetic fields. We explore the geometric phase in the Kerr and Schwarzschild geometries, which have significant astrophysical implications. Nevertheless, the work can be extended to any spacetime background including that of time-dependent. In the Kerr background, i.e. around a rotating black hole, geometric phase reveals both the Aharonov-Bohm effect and Pancharatnam-Berry phase. However, in a Schwarzschild geometry, i.e. around a nonrotating black hole, only the latter emerges. We expect that our assertions can be validated in both the strong gravity scenarios, like the spacetime around black holes, and weak gravity environment around Earth.
\end{abstract}

\keywords{geometric phase; Dirac equation; semi-classical theory; classical black holes; field theory; curved spacetime.}

\ccode{PACS numbers: 03.65.Vf, 03.65.Pm, 03.65.Sq, 04.70.Bw, 04.62.+v}


\section{Introduction}
\label{sec:intro}

One of the most fascinating features in general relativity (GR) is its 
geometric nature. Hence, all the conventional physics involved with 
geometric structure, whether in classical and quantum regimes, are expected
to reveal and also very much be suited in GR (see, e.g., Ref.\refcite{Thorne1}, where entire classical physics has been explored in geometric approaches).
In this connection, two classic widely explored quantum geometric features: 
Aharonov-Bohm (AB) effect \cite{AB1} and Pancharatnam-Berry (PB) \cite{pancharatnam1, Berry1} phase, come in question. These geometric effects are mostly explored in the presence of magnetic field in the system. While the AB effect does not necessarily require a varying field, the PB phase needs it. Indeed the magnetic field also exhibits geometric character on its own, which further triggers geometric phase onto, e.g., a spinor propagating in it. 

There are certain synonymity between magnetic and gravitational effects.
For example, both triggers curvature in the underlying spacetime; 
exhibit their respective emission mechanisms (electromagnetic radiation 
in the former and gravitational radiation, which however only emerges 
from its quadrupolar nature, in the latter) propagating with the speed 
of light; produce splitting in energy of spin-half particles (see, e.g., Refs.
\refcite{Parker1,book}), just to mention a few. Naturally, the above 
mentioned quantum effects arose in the magnetic field are expected to be 
revealed in the presence of gravitational field as well. Such 
explorations were attempted in local coordinates 
earlier \cite{Mukhopadhyay1,book,Dixit1}. However, their global exploration is very limited in the literature, to the best of our knowledge.
On the other hand, there are many astrophysical and cosmological phenomena involved with spinors, e.g. baryogenesis, neutrino emissions 
from active galactic nuclei (AGNs) and their possible oscillation, neutrino
dominated accretion disks, compact objects with Fermi degenerate gas etc., where investigation of global neutrino, in general spinor, propagation and evolution is important \cite{Mukhopadhyay2,Mukhopadhyay3,
Mohanty1,Kosteleck1,Piriz1,Chen1,Cardall1,Das1,Cook1}. Nevertheless, all the features are involved with quantum mechanics (QM) in the classical gravitational background, hence semi-classical effects.

Reconciling the two main foundations of modern physics, i.e. GR and QM, is one of the biggest challenges for both physicists and mathematicians. Semi-classical formalism has been a competent tool to study the behavior of quantum particles in classical gravitational (also electromagnetic) fields. Indeed, dynamics of Dirac fermions in a gravitational background has 
been studied for long time and found to have uniqueness and Hermiticity 
problem in the Hamiltonian formalism. Different courses of action 
\cite{Parker1,Huang1,Obukhov1,Obukhov2} were taken by different authors 
to resolve these issues. Earlier authors \cite{Gorbatenko1,Gorbatenko2} 
showed that the pseudo-Hermitian QM approach is equivalent to the approach 
taken by other authors \cite{Parker1,Huang1}, where the 
latter group used a relativistically invariant scalar product (Parker scalar product) for the state vectors. These both approaches are `standard' processes in non-Hermitian biorthogonal QM (see Ref. \refcite{Brody_2013} for review). In this paper, we mainly focus on the appearance of geometric phase in the form of AB effect or PB phase in the case of spinors traversing in gravitational field and will not elaborately look into the non-hermiticity issue of the Hamiltonian.

Berry \cite{Berry1}, in a seminal paper, showed that for a quantum system, an eigenstate of  an instantaneous Hamiltonian will return to its initial state if the Hamiltonian returns to its initial state, but only acquiring an extra phase factor known as `Berry phase'. In this case, it was assumed that the change of environment is a slow or adiabatic process. However, later on, it was proved that this geometric phases can occur in a much more general settings \cite{Samuel1,Aharonov1}.  

In this work, we study geometric effects of Dirac particles in a gravitational background. For this purpose, we deal with the 
Dirac Hamiltonian under gravitational fields 
in the $\eta$--representation, often called pseudo-Hermitian approach, 
and find the geometric phase for both the Kerr and Schwarzschild geometries. We further recognize the differences in effects appeared between these two metrices. While the Kerr metric exhibits the AB effect, in addition to PB phase, the Schwarzschild metric does not. 

In section 
\ref{Pseudo-H}, we recall the QM in the $\eta$--representation. Subsequently, in section \ref{Dirac-H} we derive the Dirac Hamiltonian for the Kerr and Schwarzschild metrices in the $\eta$--representation.
We derive the Dirac Hamiltonian in the non-relativistic approximation for slowly moving particles in section \ref{NRlimit}. 
Further, section \ref{BP} comprises the study of geometric phase for a  
Dirac particle in these two different gravitational backgrounds. Finally,
in section \ref{Conclusion}, we summarize and conclude our results.

\section{Quantum mechanics in the $\eta$--representation}
\label{Pseudo-H}
For completeness, we describe briefly the formalism of QM in the $\eta$--representation \cite{Gorbatenko1,Gorbatenko2,Bender1,Mostafazadeh1} in this section. If there exists an invertible operator $\rho$ which satisfies 
\begin{equation}
\rho H \rho^{-1} = H^{\dagger}
\end{equation}
and an operator $\eta$, which satisfies 
\begin{equation}
\rho = \eta^{\dagger} \eta,
\label{eta1}
\end{equation}
then in the $\eta$--representation, the self-conjugate Hamiltonian turns out to be 
\begin{equation}
H_{\eta} = \eta H \eta^{-1} = H_{\eta}^{\dagger}.
\end{equation}
The wave function in the $\eta$--representation is related to the wave function in the initial representation as
\begin{equation}
\Psi = \eta \psi,
\label{eta2}
\end{equation}
where the wave functions satisfy the following Schr\"{o}dinger's equations
\begin{equation}
i \frac{\partial \psi}{\partial t} = H \psi,
\label{Schro_eq1}
\end{equation}

\begin{equation}
i \frac{\partial \Psi}{\partial t} = H_{\eta} \Psi.
\label{Schro_eq2}
\end{equation}
The scalar product in initial representation is
\begin{equation}
(\phi, \psi)_{\rho} = \int d^3x (\phi^{\dagger} \rho \psi),
\label{scalar_product1}
\end{equation}
which is the same as the Parker scalar product \cite{Parker1,Huang1}, 
whereas in the $\eta$--representation the scalar product takes the form 
that of standard Hermitian QM in flat space, given by

\begin{equation}
(\Phi, \Psi) = \int(\Phi^{\dagger} \Psi)d^3x.
\label{scalar_product2}
\end{equation}
From equations (\ref{eta1}) and (\ref{eta2}), we can see that
\begin{equation}
(\phi, \psi)_{\rho} = (\Phi, \Psi).
\end{equation}

\section{Dirac Hamiltonian in the Kerr and Schwarzschild metrices}
\label{Dirac-H}

Throughout the paper, we use $\hbar=G=c=1$ unless mentioned otherwise.
The Dirac Lagrangian in curved Riemann manifold is 
\begin{equation}
L =  i \bar{\psi} \gamma^{\mu} D_{\mu} \psi - m \bar{\psi} \psi,
\label{Lagrangian}
\end{equation}
where the adjoint spinor $\bar{\psi} = \psi^\dagger \gamma^0 $ and the covariant derivative is defined as $D_{\mu} = \partial_{\mu}+\Gamma_{\mu}$, with $\Gamma_{\mu}$ being the spinorial affine connection \cite{Parker1,Huang1}. Here $m$ is the mass of the Dirac particle, $\psi$ is the four-component column bispinor and $\gamma^{\alpha}$-s are the $4 \times 4$ Dirac matrices satisfying 

\begin{equation}
\gamma^{\alpha}\gamma^{\beta}+\gamma^{\beta}\gamma^{\alpha} =  2 g^{\alpha \beta} I_4,
\end{equation}
where $I_4$ is the $4 \times 4$ identity matrix and $g^{\alpha\beta}$ is the contravariant metric tensor of curved spacetime.

The Lorentz invariant action is 
\begin{equation}
S =  \int d^4x \sqrt{-g} L,
\end{equation}
where $g={\rm det}(g_{\mu\nu})$.

The Dirac equation in curved spacetime is straightforward to derive using Euler-Lagrange formalism for the above Lagrangian given by equation (\ref{Lagrangian}), treating $\psi$ and $\bar{\psi}$ as independent variables, which turns out to be 

\begin{equation}
(i \gamma^{\mu} D_{\mu} - m) \psi (x) = 0.
\label{Dirac equation1}
\end{equation}
The global gamma matrices ($\gamma^{\alpha}$) are related to the local gamma matrices ($\gamma^{a}$) by the relation 

\begin{equation}
\gamma^{\alpha} = e^{\alpha}_{a} \gamma^{a},
\label{tetrad_relation}
\end{equation}
where $e^{\alpha}_{a}$-s are the tetrads which are defined by 

\begin{equation}
g_{\mu \nu} = e^{a}_{\mu} e^{b}_{\nu} \eta_{ab}.
\end{equation}
By our chosen convention,

\begin{equation}
\eta_{ab} = diag[1,-1,-1,-1].
\end{equation}
We reduce the Dirac equation given by equation (\ref{Dirac equation1}) in the Schr\"{o}dinger form to obtain the global Hamiltonian operator from equation (\ref{Schro_eq1}), which turns out to be

\begin{eqnarray}
H = - i \Gamma_{t}- i (g^{tt})^{-1}\gamma^{t}  \left[\gamma^r (\partial_r +\Gamma_r)\right. \nonumber \\
\left. +\gamma^{\theta} (\partial_{\theta} + \Gamma_{\theta}) +\gamma^{\phi} (\partial_{\phi} + \Gamma_{\phi})\right] + (g^{tt})^{-1} \gamma^t m. 
\end{eqnarray}
This Hamiltonian is self-adjoint under the scalar product defined in equation (\ref{scalar_product1}) where $\rho=\sqrt{g^{tt}}\gamma^t \gamma^0$. However, we will obtain the Hamiltonian in the $\eta$-representation to work with the formalism used for the standard flat Hilbert space. In the following subsections, we will find the Hamiltonians in the $\eta$-representation in both the Kerr and Schwarzschild metrices.

\subsection{Kerr metric}
\label{Kerr}

The Kerr metric in the Boyer-Lindquist coordinates is

\begin{align}
ds^2 = \left(1-\frac{2Mr}{\rho^2}\right)dt^2 + \frac{4Mra \sin^2 \theta}{\rho^2} dt d\phi - \frac{\rho^2}{\Delta} dr^2 \nonumber \\
 - \rho^2 d\theta^2 - \left[(r^2+a^2)\sin^2\theta + \frac{2Mra^2 \sin^4\theta}{\rho^2}\right] d\phi^2 
\label{kerr_metric}
\end{align}
where $\rho^2 = r^2+a^2 \cos^2\theta$ , $\Delta = r^2-2Mr+a^2$, $M$ is the mass
of black hole and $a$ is the Kerr parameter. Without any loss of generality, we choose the Schwinger gauge of tetrad \cite{Schwinger1,Gorbatenko3,Neznamov1}, given by

\begin{eqnarray}
e^t_0=\sqrt{g^{tt}} , e^r_1 = \frac{\sqrt{\Delta}}{\rho} , e^\theta_2 = \frac{1}{\rho} , \nonumber \\ e^\phi_3 = \frac{1}{\sin\theta \sqrt{\Delta}\sqrt{g^{tt}}} , e^\phi_0 = \frac{2Mar}{\rho^2\Delta\sqrt{g^{tt}}}.
\end{eqnarray}
The resultant Hamiltonian in the $\eta$-formalism turns out to be \cite{Gorbatenko4}

\begin{align}
H_\eta = \frac{m}{\sqrt{g^{tt}}} \gamma^0 - i \frac{\sqrt{\Delta}}{\rho \sqrt{g^{tt}}} (\frac{\partial}{\partial r} + \frac{1}{r}) \gamma^0 \gamma^1 \nonumber \\
- i \frac{1}{\rho \sqrt{g^{tt}}} (\frac{\partial}{\partial \theta}+\frac{1}{2} \cot\theta) \gamma^0\gamma^2 -i \frac{1}{g^{tt}\sqrt{\Delta}\sin\theta}\frac{\partial}{\partial \phi} \gamma^0\gamma^3 \nonumber \\
- i \frac{2Mar}{g^{tt}\rho^2\Delta}\frac{\partial}{\partial \phi} - i \frac{1}{2}\frac{\partial}{\partial r}(\frac{\sqrt\Delta}{\rho\sqrt{g^{tt}}}) \gamma^0 \gamma^1 -i \frac{1}{2}\frac{\partial}{\partial \theta}(\frac{1}{\rho\sqrt{g^{tt}}}) \gamma^0 \gamma^2 \nonumber \\
+ i \frac{{\sqrt{g^{tt}} \Delta M a \sin \theta}}{2 \rho} (\frac{\partial}{\partial r} (\frac{r}{g^{tt}\rho^2\Delta}) \gamma^3 \gamma^1 \nonumber \\
+\frac{1}{\sqrt{\Delta}} \frac{\partial}{\partial \theta} (\frac{r}{g^{tt}\rho^2\Delta}) \gamma^3 \gamma^2).
\label{Hk}
\end{align}
Note that $\gamma^3\gamma^2$ and $\gamma^3\gamma^1$ can be 
respectively written as $i\gamma^0\gamma^1\gamma^5$ and $-i\gamma^0\gamma^2\gamma^5$. Therefore, by rearranging equation (\ref{Hk}), we obtain the Hamiltonian in a compact form as

\begin{eqnarray}
H_\eta = ({\sqrt{g^{tt}}})^{-1}(\gamma^0 m + \gamma^0 \gamma^j (p_j-i A_j) \nonumber \\
+ i\gamma^0\gamma^j\gamma^5 k_j + e^{\phi}_{0} p_{\phi}),
\label{hamil_kerr}
\end{eqnarray}
where $A_1= \frac{\sqrt{\Delta}}{\rho r}+\frac{\sqrt{g^{tt}}}{2}\frac{\partial}{\partial r}\left(\frac{\sqrt{\Delta}}{\rho \sqrt{g^{tt}}}\right)$, $A_2 = \frac{\cot \theta}{2 \rho}+\frac{\sqrt{g^{tt}}}{2}\frac{\partial}{\partial \theta}\left(\frac{1}{\rho \sqrt{g^{tt}}}\right)$ and $A_3 = 0$; 
$k_1 =  \frac{ig^{tt}M a \sqrt{\Delta} \sin\theta}{2 \rho} \frac{\partial}{\partial \theta} \left(\frac{r}{g^{tt}\rho^2\Delta}\right)$, $k_2 =  -\frac{i g^{tt}M a \Delta \sin\theta}{2 \rho} \frac{\partial}{\partial r}\left(\frac{r}{g^{tt}\rho^2 \Delta}\right)$ and $k_3=0$.

Interestingly, the Hamiltonian is involved with a vector
and a pseudo-vector terms, while the latter vanishes in the Schwarzschild
geometry.
As $H_\eta = i \frac{\partial}{\partial t}= p_t$ and $ \partial_0 = e^t_0 \partial_t + e^{\phi}_0 \partial_\phi$, equation (\ref{hamil_kerr}) reduces to

\begin{eqnarray}
\left[p_0 - \gamma^0\{m + \gamma^j (p_j-i A_j)\right. \nonumber \\
\left. + i\gamma^j \gamma^5 k_j\}\right] \Psi = 0.
\label{hamil_kerr1}
\end{eqnarray}
It is important to note that we use the tetrad transformation relations to construct $4$-momentum in local spacetime as 
\begin{equation}
p_j = e^{\mu}_j p_{\mu},
\end{equation}
where $\mu$ (any Greek index) represents global coordinates ($t,r,\theta,\phi$) and $j$ (any roman index) represents flat coordinates ($0,1,2,3$),  
$p_\mu = i (\partial_t,-\partial_r,-\partial_\theta,-\partial_\phi)$ and 
$p_j = i (\partial_0,-\partial_1,-\partial_2,-\partial_3)$. Here $A_j (= A_1,A_2,A_3)$ is analogous to the magnetic vector potential and 
$k_j(=k_1,k_2,k_3)$ is a ``pseudo-vector" potential, which is purely a 
property of the Kerr metric. We however name both of them in general as 
``gravito-magnetic potential" in the Kerr geometry. The appearance of pseudo-vector potential $k_j$ is an interesting feature in the Kerr geometry which arises due to the chirality in the system due to the rotation of spacetime. The ``gravito-magnetic potentials" are generally functions of spacetime coordinate, namely $r, \theta$ in the case of Kerr metric.

\subsection{Schwarzschild metric}
\label{Schwarzschild}
We obtain the Schwarzschild metric from equation (\ref{kerr_metric}) with $a=0$, given by

\begin{align}
ds^2 = \left(1-\frac{2M}{r}\right)dt^2 - \left(1-\frac{2M}{r}\right)^{-1} dr^2 \nonumber \\
- r^2 d\theta^2 - r^2\sin^2\theta  d\phi^2
\end{align}
with tetrad choice as

\begin{align}
e^t_0=\sqrt{g^{tt}} , e^r_1 = \sqrt{g^{rr}}=(\sqrt{g^{tt}})^{-1} , e^\theta_2 = \frac{1}{r} , \nonumber \\ e^\phi_3 = \frac{1}{r \sin\theta}. 
\end{align}
The Hamiltonian in the $\eta$-representation is

\begin{align}
H_\eta = \frac{m}{\sqrt{g^{tt}}} \gamma^0 - i \frac{\sqrt{1-\frac{2M}{r}}}{\sqrt{g^{tt}}} (\frac{\partial}{\partial r} + \frac{1}{r}) \gamma^0 \gamma^1 \nonumber \\
- i \frac{1}{r \sqrt{g^{tt}}} (\frac{\partial}{\partial \theta}+\frac{1}{2} \cot\theta) \gamma^0\gamma^2 -i \frac{1}{{\sqrt{g^{tt}}} r \sin\theta}\frac{\partial}{\partial \phi} \gamma^0\gamma^3 \nonumber \\
- i \frac{1}{2}\frac{\partial}{\partial r}(\frac{\sqrt{1-\frac{2M}{r}}}{\sqrt{g^{tt}}}) \gamma^0 \gamma^1
\end{align}
By rearranging we obtain
\begin{align}
H_\eta = ({\sqrt{g^{tt}}})^{-1}(\gamma^0 m + \gamma^0 \gamma^j (p_j-i A^s_j))
\label{hamil_schwarz}
 \end{align}
where $A^s_1 = \frac{1}{r}\sqrt{1-\frac{2M}{r}} + \frac{\sqrt{g^{tt}}}{2}\frac{\partial}{\partial r} (\frac{\sqrt{1-\frac{2M}{r}}}{\sqrt{g^{tt}}})$ , $A^s_2 = \frac{\cot\theta}{2r}$ and $A^s_3 = 0$.

Following the similar procedure as in equation (\ref{hamil_kerr1}), for the 
Schwarzschild metric, we can write,

\begin{align}
\left[p_0 -\gamma^0\{m + \gamma^j (p_j-i A^s_j)\}\right] \Psi = 0.
\label{hamil_scharz1}
\end{align}
Here $A^s_j$ is the ``gravito-magnetic potential" in the Schwarzschild geometry. Similar to the case of Kerr metric, as is evident that the ``gravito-magnetic potentials" are functions of spatial coordinates.
It is important to note that equations (\ref{hamil_kerr1}) and (\ref{hamil_scharz1}) are the Dirac equations corresponding to the Kerr and Schwarzschild backgrounds respectively written in a compact form.

\section{Non-Relativistic approximation of Dirac Hamiltonian}
\label{NRlimit}

We chose Dirac representation for further calculation where 

\begin{equation}
\gamma^0 =
\begin{pmatrix}
I_2 & 0 \\
0 & I_2
\end{pmatrix},
\end{equation}

\begin{equation}
\gamma^i =
\begin{pmatrix}
0 & \sigma^i \\
-\sigma^i & 0
\end{pmatrix},
\end{equation}
where $I_2$ is a $2 \times 2$ unit matrix and $\sigma^i$ are the Pauli spin matrices.

 From equation (\ref{hamil_kerr1}), we write the Dirac equation as
 \begin{equation}
  \begin{pmatrix}
  -p_0+i  \vec{\sigma} \cdot \vec{k}  +m & \vec{\sigma}\cdot\vec{\Pi}_A \\
     \vec{\sigma}\cdot\vec{\Pi}_A &-p_0+i  \vec{\sigma} \cdot \vec{k}  -m
  \end{pmatrix} \Psi = 0,
 \end{equation}
where
\begin{eqnarray}
\vec{k}=(k_1,k_2,0), \nonumber \\
\vec{\Pi}_A=\vec{p}-i\vec{A}.
\end{eqnarray}
It is straightforward to derive the non-relativistic limit of the Hamiltonian following the standard method \cite{sakurai1967advanced}.
Since $\vec{A}$ and $\vec{k}$ are time independent (the metric is time independent), the time dependence of $\Psi$ is given by
\begin{equation}
\Psi = \Psi(\vec{x},t)|_{t=0} e^{-iEt}.
\end{equation}
Here $E$ is the eigenvalue of the operator $p_0= i\partial_0$, with 
$\Psi$ being the eigenfunction, given by
\begin{eqnarray}
\Psi = \begin{pmatrix}
\Psi_A \\
\Psi_B
\end{pmatrix},
\end{eqnarray}
 and the coupled equations can be written as

\begin{subequations}
\begin{align}
(\vec{\sigma}\cdot \vec{\Pi}_A) \Psi_B = (E-i\vec{\sigma} \cdot \vec{k} - m) \Psi_A,  \label{eq:subeq1} \\
(\vec{\sigma}\cdot \vec{\Pi}_A) \Psi_A = (E-i\vec{\sigma} \cdot \vec{k} + m) \Psi_B. \label{eq:subeq2}
\end{align}
\end{subequations}
Substituting $\Psi_B$ from equation \eqref{eq:subeq2} in equation \eqref{eq:subeq1}, we obtain
\begin{align}
(\vec{\sigma}\cdot \vec{\Pi}_A)\frac{1}{(E-i\vec{\sigma} \cdot \vec{k} + m)}(\vec{\sigma}\cdot \vec{\Pi}_A) \Psi_A = (E-i\vec{\sigma} \cdot \vec{k} - m) \Psi_A
\label{coupled_eq_sub}
\end{align} 

Now assuming a slowly rotating spacetime (small Kerr parameter) and the particles traveling with low velocity $v << c$,
\begin{equation}
E \sim m, \hspace{0.5cm} |\vec{k}|<<m 
\end{equation}
and also defining non-relativistic energy 
\begin{equation}
E^{NR} = E - m,
\end{equation}
We can expand the term 
\begin{eqnarray}
\frac{1}{(E-i\vec{\sigma} \cdot \vec{k} + m)} = \frac{1}{2m}\left(\frac{2m}{E^{NR}+2m-i\vec{\sigma}\cdot\vec{k}}\right) \nonumber\\
=\frac{1}{2m}\left(1-\frac{E^{NR}-i\vec{\sigma}\cdot\vec{k}}{2m}+.....\right).
\end{eqnarray}
Keeping only the leading order term, we can write from equation (\ref{coupled_eq_sub}), 
\begin{eqnarray}
\frac{1}{2m} (\vec{\sigma}\cdot\vec{\Pi}_A)(\vec{\sigma}\cdot\vec{\Pi}_A) \Psi_A = (E^{NR} - i\vec{\sigma}\cdot\vec{k}) \Psi_A,
\end{eqnarray}
which becomes, after some algebra,

\begin{eqnarray}
\left[\frac{\Pi^2}{2m} + \sigma \cdot (i\vec{k}-\frac{i}{2m}\vec{B}_g)\right] \Psi_A = E^{NR} \Psi_A,
\label{Hamil_NR_kerr1}
\end{eqnarray}
where $\vec{B}_g=\vec{\nabla}\times\vec{A}$, as defined earlier, is the 
effective gravito-magnetic field.
Using the Schr\"odinger equation
\begin{equation}
H^{NR} \Psi_A = E^{NR} \Psi_A,
\end{equation}
we can write
\begin{equation}
H^{NR} \Psi_A = i \partial_0 \Psi_A =\left[\frac{\Pi^2}{2m} + \vec{\sigma} \cdot \vec{B}^{kerr}_g\right] \Psi_A, 
\label{Hamil_NR_Kerr2}
\end{equation}
where $\vec{B}^{kerr}_g = (i\vec{k}-\frac{i}{2m}\vec{B}_g)$, what involves
with a gravitomagnetic interaction due to the Kerr metric, which includes a magnetic field analogue term $\vec{B}_g$ and a vector potential analogue term $\vec{k}$. Importantly note that in this case, $\vec{B}^{kerr}_g $ is a function of spatial coordinates and a general scenario is being treated,
on the contrary to the local exploration performed earlier \cite{Mukhopadhyay1}. It is also important to note that $\Psi_A$ is the Schr\"{o}dinger-Pauli $2$-component spinor in non-relativistic QM multiplied by $e^{-imt}$ unlike $\Psi$, which is a $4$-component Dirac bispinor.

Similarly for the Schwarzschild case, from equation (\ref{hamil_scharz1}), we can write
\begin{equation}
  \begin{pmatrix}
  -p_0+m & \vec{\sigma}\cdot\vec{\Pi}^{s}_A \\
     \vec{\sigma}\cdot\vec{\Pi}^{s}_A & -p_0-m
  \end{pmatrix} \Psi = 0  
 \end{equation}
and the corresponding non-relativistic Hamiltonian can be written as
\begin{equation}
{H^s}^{NR} \Psi_A  =\left[\frac{{\Pi^s}^2}{2m} + \vec{\sigma} \cdot \vec{B}^{sch}_g\right] \Psi_A,
\label{Hamil_NR_Sch1}
\end{equation}
where $\vec{B}^{sch}_g = -\frac{i}{2m}\vec{B}^{s}_g$. Similar to the case of Kerr metric, $\vec{B}^{sch}_g $ is also a function of spatial coordinates and based on it a general scenario is being treated below in section \ref{BP}.

\section{Geometric Phase}
\label{BP}
 
To find the geometric phase, we construct parameter space with a Poincar\'e sphere choosing the vectors $\vec{B}^{kerr}_g$ and $\vec{B}^{sch}_g$ for the Kerr and Schwarzschild geometries respectively, just like the case of Dirac particle traveling in a magnetic field. Henceforth, we write these two effective gravitational interaction terms $\vec{B}^{kerr}_g$ and $\vec{B}^{sch}_g$  as $\vec{B}$ in general.

Therefore, we can write the interaction term between particle's spin and background spacetime curvature of the Hamiltonian from equations (\ref{Hamil_NR_Kerr2}) and 
(\ref{Hamil_NR_Sch1}) as \begin{equation}
H_{int}=  \vec{\sigma} \cdot \vec{B}.
\end{equation}

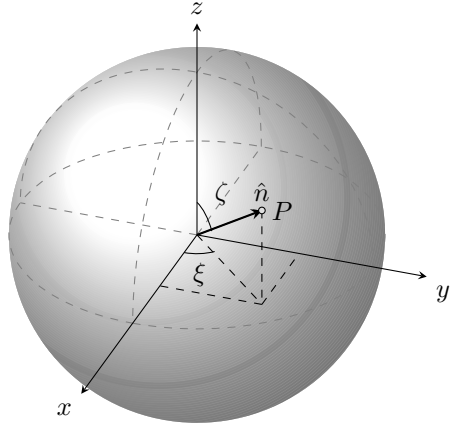
\begin{figure}
\centering
\tdplotsetmaincoords{60}{110}
\pgfmathsetmacro{\rvec}{.8}
\pgfmathsetmacro{\thetavec}{57}
\pgfmathsetmacro{\phivec}{48}
\begin{tikzpicture}[tdplot_main_coords, scale = 2.5]
\coordinate (P) at ({1/sqrt(3)},{1/sqrt(3)},{1/sqrt(3)});
 
\shade[ball color = white, opacity = 0.5] (0,0,0) circle (1cm);
 
\tdplotsetrotatedcoords{0}{0}{0};
\draw[dashed, tdplot_rotated_coords, gray] (0,0,0) circle (1);
 
\tdplotsetrotatedcoords{90}{90}{90};
\draw[dashed, tdplot_rotated_coords, gray] (1,0,0) arc (0:180:1);
 
\tdplotsetrotatedcoords{0}{90}{90};
\draw[dashed, tdplot_rotated_coords, gray] (1,0,0) arc (0:180:1);
 
\draw[dashed, gray] (0,0,0) -- (-1,0,0);
\draw[dashed, gray] (0,0,0) -- (0,-1,0);
 
\draw[-stealth] (0,0,0) -- (1.80,0,0) node[below left] {$x$};
\draw[-stealth] (0,0,0) -- (0,1.30,0) node[below right] {$y$};
\draw[-stealth] (0,0,0) -- (0,0,1.30) node[above] {$z$};
\draw[thick, -stealth] (0,0,0) -- (P) node[right] {$P$};
\draw[-stealth] (0,0,0) -- (P) node[above] {$\hat{n}$};

\draw[thin, dashed] (P) --++ (0,0,{-1/sqrt(3)});
\draw[thin, dashed] ({1/sqrt(3)},{1/sqrt(3)},0) --++
(0,{-1/sqrt(3)},0);
\draw[thin, dashed] ({1/sqrt(3)},{1/sqrt(3)},0) --++
({-1/sqrt(3)},0,0);
\draw[thin, dashed] (0,0,0) --++
({1/sqrt(3)},{1/sqrt(3)},0);

\tdplotdrawarc{(0,0,0)}{0.2}{0}{\phivec}{anchor=north}{$\xi$}
    \tdplotsetthetaplanecoords{\phivec}
    \tdplotdrawarc[tdplot_rotated_coords]{(0,0,0)}{0.2}{0}%
        {\thetavec}{anchor=south west}{$\zeta$}
 
\draw[fill = lightgray!50] (P) circle (0.5pt);

\end{tikzpicture}

\caption{Parameter space defined by the Poincar\'e sphere with $\vec{B}$ in $\hat{n}$ direction.}
\label{fig:Parameter space}
\end{figure}
We expand this matrix in a $2 \times 2$ form as
\begin{equation}
H_{int}= |\vec{B}|
  \begin{pmatrix}
     \cos\zeta &\sin\zeta \exp(-i\xi) \\
     \sin\zeta \exp(+i\xi) & -\cos\zeta 
  \end{pmatrix},
\end{equation}
where $\zeta$ and $\xi$ are the latitude and azimuthal angles of spherical polar coordinates respectively of the parameter space constructed by the vector $\vec{B}$. Here $\rho$ is the radial coordinate in this system.
See Figure \ref{fig:Parameter space}.

To find the geometric phase, we consider one of the eigenstates of the Hamiltonian, which is

\begin{equation}
\ket{\Psi}= 
  \begin{pmatrix}
     -\sin(\frac{\zeta}{2}) \exp(-i\xi)  \\
    \cos(\frac{\zeta}{2})\\
  \end{pmatrix}.
\end{equation}
The phase is defined by
  
\begin{align}
 \Phi_B = \int_R i \bra{\Psi} (\hat{\rho}  \frac{\partial}{\partial \rho}  + \hat{\zeta}  \frac{\partial}{\rho \partial \zeta}  + \hat{\xi}  \frac{\partial}{\rho \sin\zeta \partial \xi} ) \ket{\Psi} \cdot d\vec{R},
\end{align}  
where $d\vec{R} = \hat{\rho}d\rho +\hat{\zeta} \rho d\zeta +\hat{\xi} \rho\sin\zeta d\xi$ and
the connection is

\begin{align}
\vec{A}_B= i \bra{\Psi} (\hat{\rho}  \frac{\partial}{\partial \rho}  + \hat{\zeta}  \frac{\partial}{\rho \partial \zeta}  + \hat{\xi}  \frac{\partial}{\rho \sin\zeta \partial \xi} ) \ket{\Psi}.
\end{align} 
Simple calculation yields that

\begin{equation}
\vec{A}_B = \frac{(1-\cos\zeta)}{2\rho\sin\zeta} \hat{\xi},
\end{equation}
and 
\begin{equation}
\Phi_B =\frac{ \tilde{\xi}}{2} (1-\cos\zeta)=\frac{\Omega}{2},
\label{equation_berry_phase}
\end{equation}
where $\tilde{\xi}$ is the total integrated azimuthal coordinate and
$\Omega$ is the integrated solid angle. Thus the geometric phase in the spacetime around a black hole turns out to be of the same known conventional form, e.g., as revealed in the magnetic field.

Hence, it is in accordance with the Chern theorem, with the Berry curvature $\vec{\nabla}\times \vec{A}_B = \frac{1}{2\rho^3} \vec{\rho} $ and Chern number $C= \frac{1}{2\pi} \int_S \vec{\nabla}\times \vec{A}_{B} \cdot d\vec{S} = \frac{\Omega}{4\pi}$.
Therefore, the geometric phase is acquired by a quantum system in gravitational background due to the spin interaction with curvature, which leads to the spin dependent particle's dynamics. 

Above result further argues that the form of geometric phase does not depend on the gravitational field strength explicitly as given by equation (\ref{equation_berry_phase}). The same form of phase would be acquired by a spinor in a much slowly rotating Earth's gravitational field, and also in a nonrotating gravitational background. 
However, $\tilde{\xi}$ and $\zeta$ and their completion are determined by the nature of background field. Hence, even in Earth's gravity, (very weak) frame dragging will arise and a spinor in principle will acquire a Kerr-like effect in its dynamics. An interesting result is that geometric phase appears in 
the Kerr spacetime as a combination of an AB effect (determined only by the background vector potential $\vec{k}$) and PB phase (determined by the curl of background vector potential: $\vec{\nabla} \times \vec{A}$), 
whereas in the Schwarzschild only the latter arises.
To understand this quantitatively, it is important to identify that $\zeta$ and $\xi$ are the effective latitude and azimuthal angles originated from a net vector field, which is a combination of the ``gravito-magnetic potential ($\vec{k}$)" and ``gravito-magnetic field ($\vec{B}_g$)" for Kerr background. For the Schwarzschild case, only such ``gravito-magnetic field ($\vec{B}^s_g$)" will generate these angles.

\subsection{Possible measurement}

Generally speaking, any quantum effect is revealed when the underlying length
scale ($l$) is of the order of or less than the corresponding de Broglie 
wavelength $\lambda = \hbar/p$  of the particle, where $p$ is the momentum 
of the particle. For the geometric effects under
discussion, $l$ is typically the radius of the gravitational body. 
Previous studies, like Refs.~\refcite{Youlin1,Colella1}, have shown the existence and validity of such semi-classical effects of gravity for massive particles.

We consider a few situations to discuss the relevance of the gravitational 
geometric effects. In the case of an electron traveling on Earth's surface, 
$l \sim 6500$ km. Therefore the velocity of the electron has to be 
$v \lesssim 1.69\times 10^{-9}$ cm/sec, which is too small to be detectable.
This is nearly a particle at rest. 
On the other hand, on a neutron star surface (with radius $\sim 10$ km), this velocity will be 
$\lesssim 10^{-6}$ cm/sec, again too small. 

For a black hole, however, $l\sim GM/c^2$, where $G$ is Newton's gravitational constant
and $c$ is the speed of light. 
Therefore for geometric effects to be 
detected $M\lesssim 15/p$ gm.
For a non-relativistic proton moving with 
velocity $v \sim 1$ km/sec in a black hole spacetime, geometric phases can be 
practically significant for $M\lesssim 10^{20}$ gm, hence only for primordial 
black holes. It is known that primordial black holes of mass $\gtrsim 10^{15}$ gm do
not yet evaporate by Hawking radiation\cite{hawking}. In other words, for a $\bar{m}$ solar mass black hole, geometric effects will be practically revealed for $p\lesssim 7.5\times10^{-33}/\bar{m}$  gm~cm/sec. Hence, for an astrophysical black hole, the required $p$ turns  out to be too tiny. However, there is no surprise in it. It is evident from the
expression for the gravito-magnetic field (as well as potential) that 
it scales as inverse square (inversely) with distance. Now the size of 
primordial black holes could be several orders of magnitude smaller than 
that of astrophysical black holes and neutron stars (as well as Earth). 
Therefore, gravito-magnetic field for the former turns out to be much
higher leading to large and detectable geometric phases. Although, 
gravito-magnetic field does not explicitly appear in the geometric phases
in their description in the Poincar\'e sphere, the above reason effectively
is responsible for controlling geometric phases in various sites. 
Nevertheless, a full covariant study, unlike the present exploration
of non-relativistic particles in a weakly rotating black hole spacetime, would
reveal the complete picture of gravitational geometric effects and how do
they depend on the mass of black holes.

The appearance of geometric phase in our study, 
although theoretically always possible, will be practically feasible to detect around, e.g.,
primordial black holes. There are missions to infer the evidence for
primordial black holes, which are expected to be originated in early universe.
One such example is, by observing specific small interference patterns within  
gamma-ray bursts by the {\it Fermi} Gamma-ray Space Telescope one could 
argue for the existence of primordial black holes. 



\section{Conclusion}
\label{Conclusion}

There are several astrophysical, cosmological, as well as laboratory systems
including those of, e.g., condensed matter and atomic physics, which are involved with spinors, e.g. electrons, protons, neutrinos, and magnetic fields. The geometric phases are well-known/expected in these contexts. The most common formulations of geometric effects are -- AB effect and PB phase. The first one originates solely due to an external potential even with zero field, whereas the latter one arises due to the spin-field 
interaction, in the case of a spinor traveling in a magnetic field \cite{Cohen1}.

As the spacetime curvature is geometric in notion, the features exhibited in the presence of magnetic field, are naturally expected to appear in background gravitational field. Here we have shown that how the geometry of black holes can influence the propagation of spinor and in fact leads to geometric phases in them. The spinors are shown to exhibit two kinds of effect: AB effect and PB phase. While in the spacetime around a rotating black hole, both the features arise, around a static black hole only PB phase emerges. The interesting manifestation is, in the Kerr geometry the AB effect in fact originates from a pseudo-vector potential, which is solely due to the rotation of the spacetime that incorporates chirality. The PB phase includes the static effects of a gravitational background.

The similar effects are apparent in any other spacetime including that of expanding universe. In the parametric space, the underlying geometric phase turns out to be the same in feature as that in magnetic fields. However, its possible measurement depends on the length scale of the change in gravitational field.
Although the results presented here are for
non-relativistic particles and are valid in the spacetime around weakly rotating black holes,
a covariant formalism of similar effect in 
a general gravitational background is expected to reveal the similar result.
We plan to undertake that work in future. Indeed, the geometric phase
occurring in the magnetic field has been shown to remain intact
in a covariant formalism \cite{Stone1}.

Interestingly, the worldline of Dirac fermion under the influence of 
gravitational field is different from the timelike geodesic generally 
discussed in GR ignoring the spin-field coupling. However, for many 
practical purposes, a classical description suffices when length scale of 
the system $(l)$ is much larger than the de Broglie wavelength 
$\lambda = \hbar/p$ of the particle. 
Nevertheless, when $l \sim \lambda$, we have to consider the spin-dependent 
dynamics of the particle. We can observe this effect in some astrophysical 
phenomena, where the phase differences due to this spin-field interaction 
between more than one particles traveling through spacetime of different 
gravitational strengths (different spacetime points) will give rise to 
some interference pattern. Plausible candidates
might be jets, accretion flows in black hole sources.

It is also important to understand that our choice of work 
in the $\eta$-representation and the choice of Schwinger 
gauge do not affect the final results. The nature of geometric effects/phases are general,
as long as the particle is non-relativistic, and will be same irrespective of the choice of our framework.

\section*{Acknowledgments}

We thank L. Andersson (AEI), S. Banerjee (IITJ), T. Das (IISc), K. Ghosh (Vivekananda College), M. Oancea (AEI), D. Roy (RRI), J. Samuel (RRI), V. M. Vyas (RRI) for useful discussions and suggestions at various stages of the work. 
Further
thanks are due to the referee whose comments have helped us to improve the 
presentation of the work.


\end{document}